
\magnification \magstep1
\rightline {DAMTP 93-24}
\leftline {Preliminary version}
\leftline {May   1993}
\vskip .7cm
\centerline {\bf Scattering of very light charged particles}
\vskip .5cm
\centerline {J C Taylor}
\vskip .2cm
\centerline {\it DAMTP, Cambridge University, Silver Street, Cambridge,UK}
\vskip .8cm
I advance arguments against the view that the Lee-Nauenberg-Kinoshita  theorem
is relevant in practice
to the scattering of charged particles as their mass tends to
zero. I also discuss the case of massive coloured particle scattering.
\vskip 1cm
Scattering cross-sections for reactions involving, in the initial state, a
charged particle of mass $m$ contain (in perturbation theory) $\hbox {ln}~m$
terms, which become large as $m\rightarrow 0$. These logarithms come from
virtual photons which are nearly parallel to the charged particle.
In 1964, following earlier work of Kinoshita [1],
Lee and Nauenberg [1] proved that by summing over an appropriate set
of initial states the $\hbox {ln}~m$ terms could be removed. Since then it
seems not to have been generally agreed whether the initial states required for
the
theorem correspond to physically realistic situations.
(For further references, see [2].)

In 1992, Contopanagos and Einhorn [2] (referred to as CE below) published
a paper which, amongst many other things, seemed to claim with some
certainty that the Lee-Nauenberg-Kinoshita (KLN) initial-state sum does
represent
physical reality. They went further and studied qualitatively the
extent of the initial-state sum necessary to represent a typical
physical, realistic situation. This has the great virtue of focusing
the argument in a concrete way.

I remain unconvinced of the physical relevance of the initial-state sums,
and, lest the very thorough CE paper should be thought to close the
argument, I write this note to emphasize the questions which, to my mind,
remain. I do not claim any complete understanding, but I hope this note
may at any rate provoke further discussion.

In QED, soft divergences are well understood in the Bloch-Nordsiek theory.
Further, the application of the KLN theorem to final-state collinear
divergences
is uncontroversial. So I concentrate on initial-state collinear divergences
in QED.
I also discuss the relevance of the KLN theorem to soft
divergences in (perturbative) QCD, where, even with massive particles in the
initial state, there are
infrared divergences (uncancelled by
Bloch-Nordsiek {\it final}-state sums) [3].

First note that the opinion of CE is not totally clear. In the second
paragraph of the paper they state `we shall show that the requisite
initial-state sum does inevitably occur in massless theories'. But in Section 4
they say: `The equality displayed in Eq. (4.1) requires a specific relative
weighting among degenerate initial states, viz., the same phase space
normalizations that apply to final states. While this relation is an
indisputable mathematical fact, it carries the paradoxical implication that
initial-state degeneracy is to be associated with a certain relative weight
between, say, an incoming single electron of definite energy and an electron
of much lower energy accompanied by a hard but nearly collinear photon.
This conflicts with the intuitive notion of an electron beam as well as the
idea that one may prepare arbitrary linear combinations of states in Hilbert
space. A complete resolution of this paradox requires a more careful
analysis of the measurement process. While we have not carried out such an
study, we believe it would would show that ...'
(This understates the paradox: the relative {\it phases} as well as
weights of the degenerate states must be right.) The general tone
of the CE paper, however, seems to be one of great confidence in physical
relevance
of the initial-state sums.

Something very strong is being claimed: that whatever the physical
situation, collider or fixed target, etc., there will be some KLN
initial-state (with some choice of parameters) which corresponds to it.

I begin by briefly summarizing CE's formulation, which has the advantage of
being rather concrete. In the spirit of CE, I will consider an `electron'
of very small mass $m$, rather than a truely massless one. This avoids
mathematical questions about Hilbert spaces.

CE construct two asymptotic Hamiltonians, $H_A(\delta)$ and $H_A(\delta')$,
where $\delta$ and $\delta'$ are each some sort of `resolution' parameter
(or set of parameters) connected with the initial and final states.
They then define two M\o ller operators
$$ \Omega^{+} = \lim_{t\rightarrow -\infty}~e^{iHt} e^{-iH_A(\delta)t},
\eqno(1)$$
$$ \Omega^{-} = \lim_{t \rightarrow +\infty}~e^{iHt} e^{-iH_A(\delta')t}.
\eqno(2)$$
(The $H_A$ are supposed to be chosen such that these limits exist.)
Lastly they define an operator
$$ S_A(\delta, \delta') = \Omega^{-}S \Omega^{+\dagger},  \eqno(3)$$
where $S$ is the ordinary Feynman-Dyson scattering operator.

Then the following claims are made about $S_A$:-

\noindent (a) The matrix elements of $S_A$ between ordinary Fock states are
insensitive to $m$ for very small $m$ (i.e. $m/E << \delta, \delta'$).

\noindent (b) These matrix elements describe realistic scattering
experiments, provided $\delta $ and $\delta'$ are chosen suitably.

\noindent (c) There is no ambiguity about the Fock states to be chosen.
For example, if we choose, instead of a single electron state, an
electon-photon state within the resolution angle $\delta$, the result is
zero.

\noindent (d) For the example of an electron beam in a collider, $\delta$
is to be identified, at least in order of magnitude, with $r/L$,
where $L$ is the distance from the final focus (FF) to the intersection
point (IP), and $r$ is a dimension of the beam spot at IP.
At LEP, for example, CE estimate that $\delta \simeq 3 \times 10^{-6}<
m/E \simeq 10^{-5}$; so that the claimed KLN cancellation is not far
from being of practical importance.

Presumably the interpretation is that the coherent superposition of Fock
states in, for example,
$$  \Omega^{+\dagger}|\hbox {one-elecron}, {\bf p}\rangle \eqno(4)$$
is automatically and inevitably generated in the machine at FF or upstream of
it.

The reason for believing this is, I suppose, that the mechanism for
producing the beam could be analysed in the same terms, using the operator
$S_A$ in (3), and then the final states of that production process
would automatically be described in a form involving the $\Omega^{\pm}$
operators, like (4).

My first objection  to this is that the beam emerging from the production
process, considering that as a scattering process of some kind, would be
an {\it out} state; and so of the form
$$ \Omega^{-\dagger}|\hbox {one-electron}, {\bf p} \rangle, \eqno(5)$$
rather than (4). To express (5) in terms of states like (4) requires
the `collinear S-operator'
$$ \sigma = \Omega^{-\dagger}\Omega^{+}, \eqno(6)$$
which decribes the scattering of one collinear state into another,
i.e. the emission and absorption of collinear photons by the electron
The operator $\sigma$ is not unity, indeed its matrix elements in
general themselves contain collinear divergences.

There is another, related, objection. The analysis has all been done in
terms of plane-wave scattering states. This is of course an idealization.
To be realistic one needs wave-packet states of some kind. For example, the
initial beam should really be described by a wave-packet, with a
transverse size of order, say, $r'$. Consider now a one-electron-one-photon
Fock state
$$ |{\bf p}-{\bf k}, {\bf k} \rangle
\eqno(7)$$
contained in (5). Again, this should really consist of wave-packets
with limited transverse extent. Certainly this would be true for
photons radiated due to the focusing magnetic field. The photon wave
packet is diverging from the electron one at an angle of the order
of the angle $\theta $ between ${\bf p}$ and ${\bf k}$, where
$\theta  <\delta$. At the IP, therefore, the centres of the two
wave-packets will be separated by a transverse distance of order
$L\alpha$ which has a range up to a maximum of order $r$.
However, the state required in (4), being the time-reverse of an
out-state, requires the two wave-packets to be converging and to
exactly overlap at IP. Anything less than exact overlap would result in
incomplete cancellation of the collinear divergences in (3).

I have tried to articulate my unease about the assumptions made in [2],
especially the argument for the relevant magnitude of $\delta$. I do not
deny that      sort of effect is to be expected when $Lm/E$ becomes
{\it microscopic}. The electron propagator probably does not have a pole
at $E=({\bf p}^2+m^2)^{1 \over 2}$, but a branch point there. An electron
propagating over a distance $L$ probably samples a length of the
branch-cut of order $1/L$. This would correspond to an opening angle
$\delta \simeq (EL)^{-1/2}$.

In an accompanying paper [4], CE analyse `evanescent' processes, such as
helicity flip of an electron emitting a collinear photon. Although the matrix
element is proportional to $m$, the phase space has a factor ${1/ m^2}$,
and so the rate appears to be finite as $m\rightarrow 0$. CE claim that
KLN initial states will cancel this effect when ${m / E}<\delta$.
My remarks above apply equally to the order of magnitude of $\delta$ here.

Finally I briefly mention a nonabelian case. Here there are soft divergences,
uncancelled by the final-state Bloch-Nordsieck mechanism, when there are
coloured particles in the initial state even if these have mass [3]. For
example, one could take $b \bar b$ reactions in a hypothetical
unconfined world. To remove these soft divergences, the coherent
initial states would have to include soft gluons moving in all directions:
collinearity with the quarks is not relevant. It is difficult to
see how the `accelerator' producing the quark beams could also
produce coherent gluons converging on the annihilation point from
{\it all} directions.

This example may not be so removed from physics. An extension
of the QCD factorization theorem     to higher-twist would require a
meaning to be given to quark cross-sections. (The problem only
appears at higher-twist, because the uncancelled soft divergences are
suppressed by a factor $E^{-2}$.)
\vskip .6cm
I am grateful to D.R.T. Jones for discussions on the subject,
but he does   not                      have any responsibility for the
shortcomings of this note.
\vskip 1cm
\centerline {\bf References}
\vskip .3cm
\noindent [1] T. Kinoshita, J. Math. Phys. {\bf 3}, 650 (1962);
T.D. Lee and  M. Nauenberg, Phys. Rev. {\bf 133}, B1549 (1964).

\noindent [2] H.F. Contopanagos and M.B Einhorn, Phys. Rev. {\bf D45},
1291 (1992).

\noindent [3] R. Doria, J. Frenkel and J.C. Taylor, Nucl. Phys. {\bf B168},
93 (1980); A. Andrasi {\it et al}, Nucl. Phys. {\bf B183}, 445 (1981);
C. Di'Lieto {\it et al}, Nucl. Phys. {\bf B183}, 223 (1981).

\noindent [4] H.F. Contopanagos and M.B Einhorn, Phys. Rev. {\bf D45},
1322 (1992).
\bye